\documentclass[11pt]{article}
\usepackage{xspace}
\usepackage{graphicx}
\usepackage{amsmath}
\usepackage{amssymb}
\usepackage{color}

\definecolor{Red}{rgb}{1,0,0}
\definecolor{Green}{rgb}{0,1,0}
\definecolor{Blue}{rgb}{0,0,1}
\definecolor{Black}{rgb}{0,0,0}

\def\beq{\begin{equation}}
\def\eeq#1{\label{#1}\end{equation}}
\def\eeqn{\end{equation}}

\def\beqa{\begin{eqnarray}}
\def\eeqa#1{\label{#1}\end{eqnarray}}
\def\eeqan{\end{eqnarray}}

\let\bar=\overbar

\def\Dslash{\not{\hbox{\kern-4pt $D$}}}
\def\dslash{\not{\hbox{\kern-2pt $\del$}}}

\def\msb{{\bar{\ssstyle M \kern -1pt S}}}

\def\Title#1{\begin{center} {\Large {\bf #1} } \end{center}}

\begin{document}

\Title{An Experimental Review of Solar Neutrinos}

\bigskip\bigskip


\begin{raggedright}  

{\it Jeanne R. Wilson\index{Wilson, J.R.},\\
School of Physics and Astronomy\\
Queen Mary University of London\\
E1 4NS London, UK}\\

\end{raggedright}
\vspace{1.cm}

{\small
\begin{flushleft}
\emph{To appear in the proceedings of the Prospects in Neutrino Physics Conference, 15 -- 17 December, 2014, held at Queen Mary University of London, UK.}
\end{flushleft}
}

\section{Introduction}
Solar neutrinos have been the subject of experimental study for almost half a century. The initial aim was ``...to see into the interior of a star and thus verify directly the hypothesis of nuclear energy generation in stars."\cite{Bahcall:1964gx}. Since then the field has advanced a long way. This article reviews what we have learnt from solar neutrinos to date, the current experimental status of the field, and the finer details of solar models and precision measurements of neutrino oscillation parameters that could be probed with this free source of extra-terrestrial particles. 

\section{Solar Neutrino Production and Detection}
Neutrinos are produced through nuclear fusion reactions in the core of the Sun. There are two series of fusion reactions, the pp-chain starting with proton-proton fusion, which dominates both the energy and $\nu$ production, and the CNO cycle that fuses the heavier elements. The $\nu_e$ spectrum extends up above 10\,MeV (${^8B} \rightarrow {^8{\rm Be^*}} + e^+ + \nu_e$) but the $\nu$ flux is dominated by pp fusion ($p + p \rightarrow {^2{\rm H}} + e^+ + \nu_e\hspace{0.3cm}(E_\nu < 0.5\,{\rm MeV})$). 

There are three main reactions for detecting $\nu$s at solar energies:
\begin{description}
\item{Elastic Scattering (ES)}: $\nu_x + e^- \rightarrow \nu_x+ e^-$. Scattering of $\nu$s off electrons in the target can occur for all active $\nu$ flavours but the cross-section for $\nu_e$ is approximately six times that of the $\nu_\mu, \nu_\tau$ due to additional scattering channels via W-boson exchange that are not available for other flavours. The kinematics of the interaction wash out the energy response such that the correlation between incoming neutrino energy and outgoing electron energy is relatively weak, but the electron is predominantly emitted in the forward direction resulting in a strong angular correlation away from the Sun.
\item{Charged Current (CC)}: $\nu_e + n \rightarrow p + e^-$. CC interactions are exclusively sensitive to $\nu_e$. However, the kinematics provide good mapping between the electron energy and the incoming $\nu$ energy, so this channel is useful for performing spectral measurements.
\item{Neutral Current (NC)}: $\nu_x + {^2{\rm H}} \rightarrow p + n + \nu_x$. This interaction was specific to the SNO experiment, which employed a heavy water target. The measured signal resulted from the capture of the emitted neutron and thus did not contain any information on the incoming $\nu$ energy or direction, but was equally sensitive to all $\nu$ flavours and therefore able to measure the full active solar $\nu$ flux.
\end{description}
All these reactions rely on the production of visible particles in the detecting medium, either charged electrons, or gammas emitted by neutron capture. The leading technologies employed for solar $\nu$ detection to date are: 1) Water Cherenkov (both H$_2$O and D$_2$O), this provides real-time measurements with good directional and energy sensitivity above a threshold of order 4--5\,MeV. 2) Radiochemical techniques that involved CC interactions on Chlorine or Gallium and rely on the periodic summation of the daughter isotopes produced to give total integrated flux measurements above somewhat lower thresholds ($\approx$0.8\,MeV and $\approx$0.3\,MeV respectively). 3) Liquid scintillator that again gives real-time measurements with low thresholds ($\approx 0.2$\,MeV, driven by backgrounds) but no directional information due to the isotropic nature of scintillation light.

\section{Oscillation Parameters - Current Status}
After many years of measurements of a low ($\le 0.5 \times$ predicted) solar $\nu$ flux by radiochemical experiments sensitive to only $\nu_e$s, the Super Kamiokande\cite{Fukuda:2001nj} and SNO\cite{Ahmad:2001an} experiments unambiguously confirmed that the explanation for this so-called solar neutrino problem (SNP) was flavour conversion of $\nu_e$s to other flavours. Combined analysis (see eg.\cite{Gonzalez-Garcia:2014bfa}) of all solar data, in combination with results from the KamLAND experiment\cite{Abe:2008aa} confirmed large mixing angle $\nu$ oscillations dominated by mixing between mass states 1 and 2 with the parameters: 
$\sin^2\theta_{12} = 0.304^{+0.013}_{-0.012} , \Delta m^2_{12}  = (7.5^{+0.19}_{-0.17})\times10^{-5} \rm{eV}^2$\cite{Gonzalez-Garcia:2014bfa}. More precise measurements of the solar parameters are motivated by a desire to pin down the mixing matrix for leptonic CP violation searches. 

Experiments with different energy thresholds observe different fractions of the predicted solar flux, which can be explained by the MSW effect\cite{Langacker:1986jv}: At high energies ($>$5MeV) interactions with electron dense matter in the Sun cause a resonance in conversion between mass states reducing the measured flux to about a third (survival probability, $P_{ee} = \sin^2\theta_{12}$), but at low energies we see vacuum averaged oscillations over the path from Sun to Earth giving about half the expected flux (survival probability $P_{ee} = 1 - \frac{1}{2}\sin^2\theta_{12}$). The relative importance of the MSW matter term and the kinematic vacuum oscillation term in the Hamiltonian can be parameterised by the quantity, $\beta$, which represents the ratio of matter to vacuum effects as given in equation~\ref{eq:mswbeta}.
\begin{equation}
\beta = \frac{2\sqrt{2} G_Fn_eE_\nu}{\Delta m^2}
\label{eq:mswbeta}
\end{equation}
 where $G_F$ is the Fermi coupling constant, $n_e$ is the electron density, $E_\nu$ is the $\nu$ energy and $\Delta m^2$ is the square mass difference. Thus, our sensitivity to $\Delta m^2$ comes through this $\beta$ term as the survival probabilities, $P_{ee}$, are only sensitive to the angle; our best understanding of $\Delta m_{12}$ to date comes from KamLAND terrestrial measurements.

\section{Precision Oscillation Measurements}
Although solar data fits well to the MSW scenario, it has not been directly observed, in fact, all solar $\nu_e$ spectral measurements to date (Borexino\cite{Bellini:2008mr},  SuperKamiokande\cite{Renshaw:2014awa}, SNO\cite{Aharmim:2009gd}) indicate a downturn in the spectrum, with respect to solar model predicted fluxes, at low energies (3.5--5\,MeV) where MSW predicts an upturn. None of these measurements are statistically significant, given the large systematic errors associated with these low energy measurements, but there is a strong desire to obtain  direct confirmation of the MSW hypothesis and to probe this resonance region for possible new physics. 
\subsection{Tests of MSW}
The dependence on $n_e$ implies a dependence on radius within the Sun due to the radial density profile. Solar models also predict that the different $\nu$ producing interactions occur at different depths within the Sun, so the strength of the MSW effect varies for different components of the solar $\nu$ flux: $^8$B $\nu$s are produced closest to the core, at highest electron density and therefore see the strongest MSW effect\cite{Bonventre:2013loa}. Therefore, if more accurate measurements of the $^8$B solar flux could be made in the 2--5\,MeV region, this could provide a powerful probe of the MSW effect. Ideally such measurements would be made through CC channels, as the kinematics and $\frac{1}{6}$th sensitivity to $\nu_\mu, \nu_\tau$ washes out the effect in the ES interaction.

An alternative approach is an accurate flux measurement of pep $\nu$s ($p + e^- + p \rightarrow {^2{\rm H}} + \nu_e$). 
This interaction is a fundamental step in the pp fusion chain and therefore the flux can be predicted with low theoretical uncertainty ($\pm$1.5\%). Furthermore, the $\nu$s are mono-energetic, produced with 1.44\,MeV, directly in the resonance region. The experimental measurement is difficult due to the challenges of statistically separating the pep ES signal from backgrounds, in particular $^{11}$C decays, produced by cosmic ray muon interactions on $^{12}$C. ($\mu + {^{12}\rm{C}} \rightarrow {^{11}\rm{C}} + n$) followed by ($^{11}\rm{C} \rightarrow {^{11}\rm{B}} + e^+ + \nu_e $) with $T_{1/2} = 29.4{\rm mins}$. 
The triple coincidence of the incoming muon and the captured neutron before the $^{11}$C positron decay allow some rejection of this background, but the efficiency of such a tag depends on the incoming muon flux and hence the depth of the experiment. The muon flux through the Borexino detector, in the Gran Sasso laboratory at a depth equivalent to $\approx$3000\,m of water, is $\approx 4300$ per day, resulting in $\approx$27 $^{11}$C decays per 100\,ton of scintillator detection medium per day. Despite this high level of background, Borexino have extracted the pep signal and produced a flux measurement with $\approx 19$\% precision: $\Phi_{pep} = (1.6 \pm 0.3 ) \times 10^8 {\rm cm}^{-2}{\rm s}^{-1}$.\cite{Collaboration:2011nga} 

A further direct probe of the MSW effect is the ``day-night" effect that should be caused by terrestrial matter enhanced oscillations. At night, the solar $\nu$s detected have passed through the Earth, whereas in the day, they haven't so a comparison of day and night fluxes or spectra should be sensitive to any matter effects. However, for the measured mixing parameters, the asymmetry is expected to be small (of order 1\%).  The predicted asymmetry is actually more significant at higher energies, but to detect a 1\% difference  high statistics and a tight control of any systematic differences between night and day is needed.  Unfortunately, the Sun provides higher statistics at lower energies where background contributions are greater and less well understood.
SuperKamiokande has seen the first hint of the terrestrial day-night effect, with their large data set above 5\,MeV. The measured asymmetry, $A_{d/n} = -3.2 \pm 1.1({\rm stat}) \pm 0.5 ({\rm syst})$ deviates by 2.7$\sigma$ from zero.\cite{Renshaw:2013dzu}

\subsection{Searches for New Physics}
New physics effects are most likely to manifest themselves in resonance regions. A number of alternative theories have been proposed that manifest as distortions to the measured solar $\nu$ flux in the 1--5\,MeV region (the MSW resonance region), which can be probed through accurate measurement of the pep flux. 
Friedland, Lunardini and Pena-Garay propose flavour changing interactions at an allowed level that modify the conversion probability for $\nu$s below 6\,MeV resulting in a significantly reduced pep flux
.\cite{Friedland:2004pp} 
Holanda and Smirnov propose that additional oscillations to a small sterile $\nu$ fraction, not yet ruled out, would also reduce the $\nu_e$ survival probability only in this energy region.\cite{deHolanda:2010am}
Gonzalez-Garcia and Maltoni proposed that if the $\nu$ mass arises from interaction with a scalar field, the accelerant, whose effective potential changes as a function of $\nu$ density, the survival probability in this energy region would again be modified.\cite{GonzalezGarcia:2007ib}

\section{Probing Solar Properties}
Solar $\nu$ flux measurements have had a strong impact on theoretical solar models already, but there remain some ambiguities.  
\subsection{Solar Metallicity}
One issue that has come to light in recent years is a discrepancy between models and data dubbed the Solar Metallicity Problem. The abundance of heavier elements in the Sun can be measured through photospheric absorption lines. Improved 3-dimensional modelling of these line shapes (affected by doppler shifts due to convection and gaseous movement in the Sun) has resulted in lower metallicity levels than previously used in solar models.  When the lower metallicity numbers are propagated through the solar models, they imply a different location of the boundary region between the radioactive and convection zones in the Sun, which in turn affects the transport of sound waves, resulting in a previous unseen discrepancy between model predictions and helioseismic data. It is not clear which component is responsible for the discrepancy, but there are doubts as to whether photospheric metallicity measurements should apply to the central regions of the Sun. 

One way to distinguish between the high, GS98, and the low, AGS09, metallicity models\cite{Serenelli:2009yc} is to obtain a direct measurement of the solar core metallicity through the CNO $\nu$ flux, which is directly proportional to these core abundances but is also strongly temperature dependent such that uncertainties in the core temperature contribute of order 15\% uncertainty to the CNO flux predictions. However, the $^8$B flux, which we know experimentally to a precision of $\approx$3\%\cite{Aharmim:2009gd}, is even more strongly temperature dependent and can thus be used as a `thermometer' to constrain CNO predicted fluxes.

The challenge in measuring the CNO flux experimentally is to separate it from decays of $^{210}$Bi, which have a very similar spectrum in the 0--1.2\,MeV range. The Borexino experiment has placed a limit of $\Phi_{CNO} < 7.7 \times 10^8 {\rm cm}^{-2}{\rm s}^{-1}$ (95\% CL)\cite{Collaboration:2011nga} and aim to improve on this with a reduced $^{210}$Bi background in phase II. It is also possible to place constraints on the contribution of $^{210}$Bi by looking at the time evolution of $^{210}$Po decay, the daughter product of $^{210}$Bi decay, which has a half-life of 138\,days. 

\subsection{Solar Luminosity}
Solar $\nu$s can also be used to test the overall energy production of the Sun. The pp fusion reaction is the primary reaction in the pp chain and produces the overwhelming majority of $\nu$s emitted from the sun. Since photons produced in the core take hundreds of thousands of years to reach the Earth's surface, measurement of the luminosity through $\nu$s also confirms that the Sun has been in thermodynamic equilibrium over at least this timescale. The only experiment to measure the pp flux independently to date is Borexino, who measured $\Phi_{pp} = (6.6 \pm 0.7) \times 10^{10} {\rm cm}^{-2}{\rm s}^{-1}$\cite{Bellini:2014uqa}. Improvements on this measurement to a 1\% accuracy would be required to perform a rigorous comparison and to give insight into solar dynamics over timescales of order $10^5$\,years. 

\section{Solar Neutrino Experiments}
The previous section discussed a range of measurements that can be performed on solar $\nu$s, but unfortunately, the experimental requirements for each are different. It is not a case of one detector fits all. For example, the pp flux measurement requires a threshold of 0.2\,MeV or lower and thus extremely well constrained low energy backgrounds, but statistics are abundant so a large detector size is not necessary. The measurements of the day-night effect and the spectral shape of $\nu_e$ survival probability do not require such a low threshold but are statistics limited and therefore require large detectors. In this section, key experimental features and potential of some current and future solar $\nu$ detectors are discussed. 

\subsection{SuperKamiokande}
The SuperKamiokande experiment is a 50\,kt water Cherenkov detector located in the Kamioka mine, Japan.\cite{Renshaw:2014awa} Since commencing operation in 1996 the experiment has amassed considerable statistics through four running periods, making this experiment the leader in statistics-sensitive day-night measurements. 
Solar $\nu$ events are detected above a threshold of $\approx$4.5\,MeV, limited by radioactive backgrounds at lower energies. Through ES interactions  the detector has good directional sensitivity of $\approx$ 25$^\circ$, limited by multiple Coulomb scattering and the 11k 20\,inch PMTs afford a position resolution of 52\,cm and energy resolution of 14\%. At a depth of 2700 metres of water equivalent (m.w.e.), the muon rate in Super-Kamiokande is 1.88\,Hz.

\subsection{Borexino}
The Borexino experiment\cite{Bellini:2008mr,Collaboration:2011nga,Bellini:2014uqa} has led the way in low energy solar measurements over the past decade with unprecedented low background levels in the 300\,t liquid scintillator target (results in the previous section).  After publishing their first solar results, they had a campaign to further purify the scintillator for phase 2, which started in 2012, in which they achieved $<0.8$\,counts/year/100\,t for Uranium  and Thorium, $20 \pm 5$\,counts/day/100\,t for $^{210}$Bi and $<5$\,counts/day/100\,t for $^{85}$Kr. Although the measurements of the various flux components (pep, $^{7}$Be, CNO) are not yet at the level of precision to distinguish between high and low metallicity solar models, further background reductions and increased statistics will result in improved measurements from Borexino in the future.

\subsection{SNO+}
SNO+ is another liquid scintillator experiment, currently under construction, in SNOLAB, Canada, with a 780\,t scintillator volume.\cite{Lozza:2014maa,Biller:2014eha}
 It is over twice the size of Borexino and, at a depth of 6000\,m.w.e., sees approximately 2 orders of magnitude less muons resulting in a much lower cosmogenic background. With target backgrounds at Borexino phase 1 levels or lower, 1 year of lifetime and a 50\% fiducial volume cut to remove external backgrounds from the solar analysis, SNO+ predicts a sensitivity of 9\% for the pep flux measurement, 7.5\% for $^8$B, 4\% and for $^7$Be. However, the SNO+ experiment's primary goal is to perform a search for neutrino-less double beta decay with $^{130}$Te isotope loaded into the scintillator volume, which excludes the solar measurements during the loading period.

\subsection{JUNO and RENO-50}
JUNO is a 20\,kT  proposed liquid scintillator experiment in development in China\cite{He:2014zwa}, designed for a mixed program including reactor $\nu$ measurements to probe the mass hierarchy, but with a 700\,m rock overburden they should also have reasonable solar $\nu$ sensitivity. The ambitious programme for scintillator and photosensor research and development is aiming for 1200 photoelectrons per MeV of light to provide an energy resolution of $3\%/\sqrt{E}$.
RENO-50\cite{Kim:2014rfa} is a similar proposed liquid scintillator reactor experiment in Korea with a 10\,kt far detector under a $\approx$900 metre of water equivalent overburden. 

\subsection{LENA}
The LENA experiment\cite{Wurm:2011zn}, proposed at the Pyhasalmi mine in Finland (4060\,m.w.e. shielding depth), is a huge, 50\,kt liquid scintillator experiment that would allow very high statistics measurements of the lower energy flux components. For example, above a 250\,keV threshold and in a 30\,kTonne fiducial volume, they could detect around 600 pep events, 600--900 CNO events (AGS -- GS model predictions),  and 8000 $^7$Be events per day. This huge $^7$Be signal provides potential to observe short time modulations of order 0.1\% amplitude and the high CNO statistics would discriminate between the high and low metallicity solar models. Given the large volume, CC reactions on the small (1\%) natural isotopic abundance of $^{13}$C also become feasible, ($\nu_e + ^{13}{\rm C} \rightarrow ^{13}{\rm N} + e^-$) though still challenging, to measure (2--3 events/day). The CC electron energy tracks the incoming $\nu$ energy with an energy threshold of just 2.2\,MeV allowing lower energy measurement of the $^8$B spectrum, and delayed decay of the $^{13}$N give some coincidence information to separate this from backgrounds.

\subsection{HyperK}
HyperKamiokande is the proposed next generation water Cherenkov experiment in Japan\cite{hyperK}. With two tanks containing 1000\,kt of H$_2$O, this detector will have a solar fiducial volume 27 times that of SuperKamiokande and would expect 200 $^8$B $\nu$ interactions above 7\,MeV per day. With these statistics it would be possible to monitor for solar core temperature variations of around 1.5\% for day to day variations and to reach a day-night asymmetry sensitivity of 0.5\% (4$\sigma$). HyperKamiokande will also measure beam $\nu$s to probe leptonic CP violation and atmospheric $\nu$s to probe the neutrino mass hierarchy.

\subsection{LENS}
The LENS experiment\cite{Raghavan:2008zz} proposes to use $^{115}$In as a target for CC interations. ($\nu_e + ^{115}{\rm In} \rightarrow e^- + 2\gamma{\rm (delayed)} + ^{115}{\rm Sn}$).
The energy threshold of only 115\,keV allows observation of 95\% of the pp spectrum. 
By using materials with two different refractive indices in a lattice LENS can effectively segment the detector to provide rejection capability for external backgrounds, and the triple coincidence from two delayed gammas aids in rejection of  the $^{115}$In beta decay background. $T_{1/2}(^{115}{\rm In}) = 6.4\times10^{14}$ years so with 10\,t of Indium, the rate of this background (endpoint 498.8\,keV) is 2.5\,MHz so a suppression factor of $10^{11}$ is required. 

\subsection{ASDC -- \textsc{Theia}}
\textsc{Theia}, is a proposed realisation of the Advanced Scintillator Detector Concept\cite{Alonso:2014fwf}: a future large water-based liquid scintillator detector, which would have the ring-imaging capability of a pure water Cherenkov detector but with the advantage of particle detection below Cherenkov threshold. This detector would have a broad physics reach, including solar $\nu$s through both ES interactions and CC, achieved by loading an isotope such as $^7$Li: $ ^7{\rm Li} + \nu_e \rightarrow ^7{\rm Be} + e^- $ with Q = 862 keV, 
although other isotopes can be considered. Low threshold CC detection such as this could allow spectral separation beyond that of any pure scintillator detector.

\section{Summary}
Our understanding of neutrino oscillations has advanced significantly since solar $\nu$s were first observed in the 1960s, but there are still a number of outstanding questions that have been summarised in this review. Solar $\nu$s provide a means to probe neutrino properties, search for new physics and probe solar structure and formation. The set of potential measurements discussed here requires a range of detection approaches and technological advances in low background, large underground detection techniques.

\bigskip
\section{Acknowledgments}
Dr. J.R. Wilson is supported by the European Union's Seventh Framework Programme, FP7/2007-2013, under the
European Research Council (ERC), grant agreement no. 278310.

\end{document}